\documentclass[twocolumn]{revtex4}%
\usepackage[paperwidth=210mm,paperheight=297mm,centering,hmargin=2cm,vmargin=2.5cm]{geometry}
\usepackage{amsfonts}
\usepackage{amsmath}
\usepackage{amssymb}
\usepackage{bm}
\usepackage{graphicx}
\usepackage{color}
\usepackage{soul}
\usepackage{epsfig}%
\usepackage[dvipsnames]{xcolor} 
  \definecolor{bleu_cite}{RGB}{0,0,255}
\usepackage[colorlinks=true,allcolors = black,citecolor=bleu_cite,  ]{hyperref} 
\usepackage{natbib}

\begin{document}
\title{Spectroscopic Signatures for the Dark Bose-Einstein Condensation of Spatially Indirect Excitons}

\author{Mussie Beian$^{1,3}$, Mathieu Alloing$^1$, Romain Anankine$^{1}$, Edmond Cambril$^2$, Carmen Gomez$^2$, Aristide Lema\^{i}tre$^2$ and Fran\c{c}ois Dubin$^{1}$} 
\affiliation{$^1$ Institut des Nanosciences de Paris, CNRS and UPMC, 4 pl. Jussieu,
75005 Paris, France}
\affiliation{$^2$ Laboratoire de Photonique et Nanostructures, LPN/CNRS, Route de
Nozay, 91460 Marcoussis, France}
\affiliation{$^3$  ICFO-Institut de Ciencies Fotonicas, The Barcelona Institute of Science and Technology, 06880 Castelldefels, Spain}

\begin{abstract}
We study semiconductor excitons confined in an electrostatic trap of a GaAs bilayer heterostructure. We evidence that optically bright excitonic states are strongly depleted while cooling to sub-Kelvin temperatures. In return, the other accessible and optically dark states become macroscopically occupied so that the overall exciton population in the trap is conserved. These combined behaviours constitute the spectroscopic signature for the mostly dark Bose-Einstein condensation of excitons, which in our experiments is restricted to a dilute regime within a narrow range of densities, below a critical temperature of about 1K.
\end{abstract}

\maketitle

Semiconductor excitons, i.e. Coulomb bound electron-hole pairs, constitute a class of composite bosons which has raised a large interest in the context of Bose-Einstein condensation (BEC). This phase transition was originally envisioned in the 1960s \cite{Moskalenko_62,Blatt_1962,Keldysh_BEC}, and fifty years of research were actually necessary to detect anticipated signatures, such as long-range spatial coherence and quantised vortices \cite{Anankine_2016}. The main reason for this unexpectedly long search was given in 2007, when Combescot et al. \cite{Monique_dark_BEC} pointed out that excitons, which exist in either optically bright or optically dark forms, depending on their total spin, always have a dark ground state. Accordingly, BEC is controlled by the macroscopic occupation of dark excitons, a conclusion which stood in striking contrast with previous experimental and theoretical research that had emphasized a condensation dominated by optically bright excitons \cite{Zim_08,Combescot_ROPP}. 

The dark nature of exciton condensation sets strong barriers to evidence the quantum phase transition. Indeed, it impedes direct measurements of the excitons momentum distribution by imaging their photoluminescence in momentum space, as for example employed with atomic gases \cite{Stringari_Book_BEC,Leggett_2006} or polaritons \cite{Ciuti_2013}.  Alternative spectroscopic techniques are thus necessary. We then note that the exciton dark-state condensation leads to a photoluminescence quenching, which is easily identified in principle since it contrasts with the classically expected increase of the optical emission as the exciton temperature is lowered \cite{Deveaud,Littlewood}. However, relating unambiguously a photoluminescence darkening to BEC is a tedious task since experimental limitations can also induce a photoluminescence bleaching, for example interactions between excitons and excess free carriers  \cite{Peyghambarian_84}, or simply non-radiative losses. 

In this work, we study two-dimensional excitons trapped at a controlled total density, i.e. including bright and dark states, kept constant while the gas is cooled down to sub-Kelvin temperatures. In the dilute regime and for a restricted range of densities only, we evidence quantitatively a photoluminescence darkening of about 30$\%$. By evaluating the strength of non-radiative channels we then show that this photoluminescence quenching reveals unambiguously the buildup of a dominant fraction of dark excitons, of around 70$\%$ below a critical temperature of around 1 Kelvin. The energy splitting between bright and dark exciton states being reduced to about 5 $\mu$eV in our heterostructure \cite{Blackwood_1998,Mashkov_97}, that is one order of magnitude less than the thermal energy at our lowest bath temperature (330mK), the dominant occupation of dark states therefore provides a quantum statistical signature for Bose-Einstein condensation.
 
\begin{figure}[h!]\label{fig1}
\centerline{\includegraphics[width=.5\textwidth]{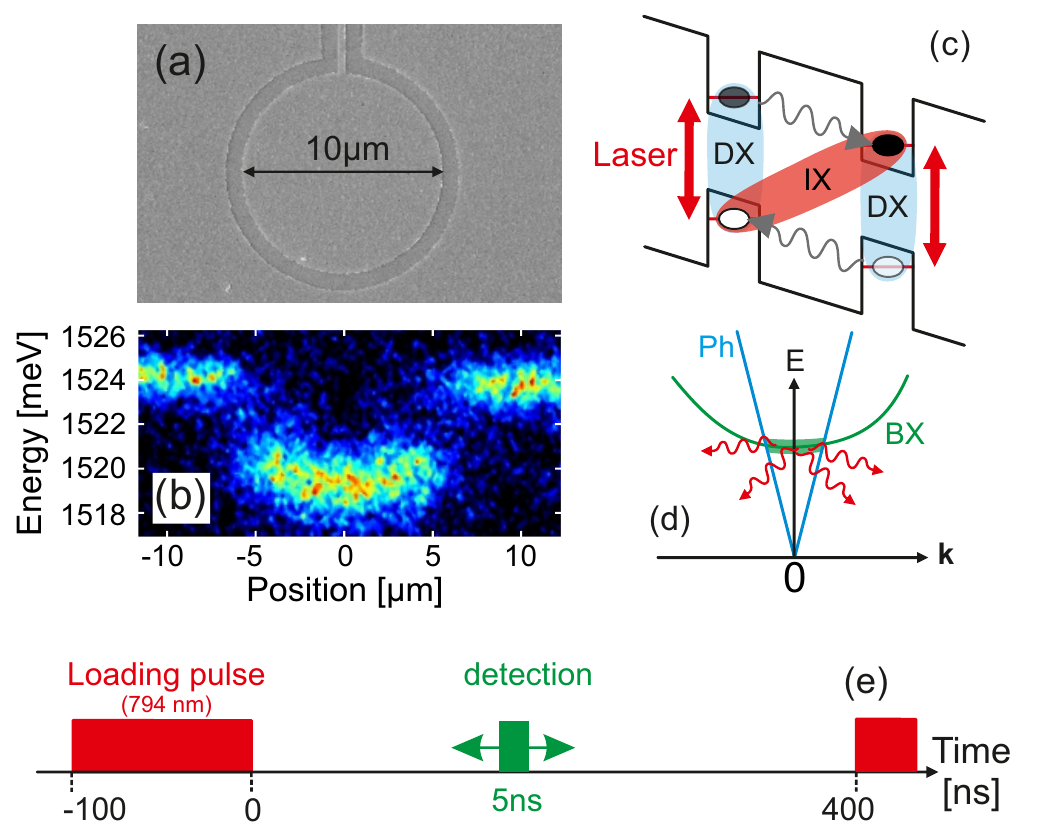}}
\caption{(a) Electron microscope image of the surface electrodes constituting the electrostatic trap. (b) Spatial profile of the confining potential realized when the center electrode is biased at (-4.8V) while the outer electrode is biased at (-3.5V) at T$_\mathrm{b}$=330 mK. The confinement profile is measured by injecting a very dilute and linearly shaped cloud of indirect excitons along the horizontal axis of the trap.  (c) Sketch of the optical injection of indirect excitons (IX), through the resonant excitation of direct excitons (DX). IXs are created once electronic carriers have thermalized towards minimum energy states (wavy grey lines). (d) The photoluminescence (wavy red arrows) is only due to the radiative recombination of lowest energy (\textbf{k}$\sim$0) bright excitons (BX), i.e. lying at an energy smaller than the intersection between excitonic (BX) and photonic (Ph) bands. (e) Our experiments rely on a 100 ns long loading pulse while the exciton dynamics is monitored in a 5 ns long and running time window which follows each laser pulse. This sequence is repeated at 2 MHz.}
\end{figure}

As illustrated in Figure 1,  we probe long-lived spatially indirect excitons \cite{Lozovik} confined in a GaAs/AlGaAs double quantum well (DQW) identical to the one probed in Ref. \cite{Anankine_2016}. The bilayer is embedded in a field-effect device, and indirect excitons are engineered by applying an electric field perpendicular to the quantum wells. This is achieved by biasing two independent and semi-transparent metallic electrodes deposited on the surface of the structure (Fig. 1.a). Minimum energy levels for electrons and holes then lie in distinct layers (Fig. 1.c), indirect excitons resulting from the Coulomb attraction between distant carriers. In the following experiments, we apply a larger potential onto the central (trap) electrode and thus create a $\sim$10 $\mu$m wide trap, indirect excitons being attracted towards the regions where the electric field is the strongest \cite{Schinner_2013,Timofeev_2013,Rapaport_2013,High_2009}. Figure 1.b shows that the trap depth is about 5 meV, and note that excitons are injected optically by a pulsed laser excitation covering the entire trap area (Fig.1.e). 

To quantify the total exciton density, i.e., including both bright and dark excitons with total "spins" equal to ($\pm$1) and ($\pm$2) respectively, we study the dynamics of the photoluminescence energy $E_\mathrm{X}$. Indeed, $E_\mathrm{X}$ is given by repulsive dipolar interactions between excitons \cite{Zim_08,Ivanov_2010,Schindler_08} and we verified experimentally that these repulsions yield a blue-shift of $E_\mathrm{X}$ scaling as $u_0 n_\mathrm{x}$ in the dilute regime (see Figure 5.b). Note that it is expected theoretically that $u_0 n_\mathrm{x}\sim$ 1meV for an exciton density $n_\mathrm{x}\sim$ 10$^{10}$ cm$^{-2}$ \cite{Ivanov_2010}. In our studies, the blueshift of the photoluminescence is actually computed as the difference between E$_\mathrm{X}$ and its value at the latest delays, i.e. when the density in the trap yields a vanishing blueshift \cite{Snoke_2009} (350 ns after termination of the laser excitation). Moreover, we analyse the dynamics of the photoluminescence integrated intensity I$_\mathrm{X}$ in order to extract the fraction of bright excitons in the trap. Indeed,  I$_\mathrm{X}$ is directly given by the product between the density of bright excitons and their optical decay rate (1/$\tau_\mathrm{opt}$). By comparing the dynamics of I$_\mathrm{X}$ and $E_\mathrm{X}$ at different bath temperatures T$_\mathrm{b}$ we then infer the occupation of dark states for any given total density in the trap.

\subsection{Electrostatic noise and inhomogeneous broadening}

Figure 2 provides an overview of the exciton spectroscopy at the lowest bath temperature (T$_\mathrm{b}$= 330 mK). First, Fig. 2.a displays the photoluminescence integrated intensity and its emission energy which decay differently, with characteristic times $\sim$60  and $\sim$120 ns respectively. This behaviour suggests that the overall exciton density decays slower than its sole bright component. Also, in Figure 2.a we note that E$_\mathrm{X}$ decreases by $\sim$ 5 meV across the delay range that we explore. Accordingly, the exciton density ranges from $\sim$5 10$^{10}$ cm$^{-2}$ at the termination of the laser pulse, to $\sim$10$^{9}$ cm$^{-2}$ 250 ns later when E$_\mathrm{X}$ becomes essentially constant. Fig. 2.b then details the dynamics of the photoluminescence spectral width $\Gamma_\mathrm{X}$. It shows that $\Gamma_\mathrm{X}$ rapidly decreases in the first $\sim$70 ns after the laser excitation, which we interpret as the manifestation for a transient photocurrent, bound to 100 pA but which nevertheless induces spectral broadening since interactions between excitons and free carriers are very strong in quantum wells \cite{Honold_1989}. By contrast, we do not resolve sizeable variations of $\Gamma_\mathrm{X}$ at longer delays ($\tau\gtrsim$ 100 ns) i.e. in the dilute regime when $n_\mathrm{x} a_\mathrm{X}^2\lesssim$0.3, $a_\mathrm{X}\sim$20 nm \cite{Mujlarov_2012} denoting the excitons Bohr radius.

\begin{figure}[h!]\label{}
\centerline{\includegraphics[width=.5\textwidth]{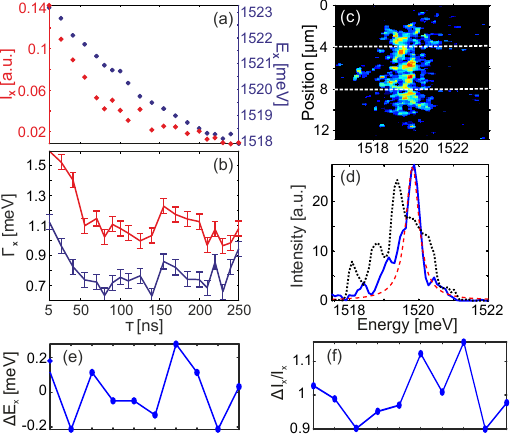}}
\caption{(a) Integrated intensity (I$_\mathrm{X}$, red points) and energy (E$_\mathrm{X}$, blue points) of the photoluminescence as a function of the delay to the termination of the laser excitation ($\tau$). (b) Dynamics of the photoluminescence spectral width ($\Gamma_\mathrm{X}$) computed by accumulating directly 10 successive spectra (red) or by evaluating a statistical average (blue). (c) Photoluminesence spectrally and spatially resolved along the diameter of the trap for a delay $\tau$=170 ns. The solid line in (d) shows the spectrum evaluated between the two dashed lines in (c). It is marked by a $\sim$400 $\mu$eV linewidth limited by the spectral resolution (dashed red line). The dotted line in (d) shows the photoluminescence at the same position for one of the successive 60 second acquisitions. (e) Variation of the photoluminescence energy $\Delta$E$_\mathrm{X}$ and variation of the normalized integrated intensity $\Delta$I$_X$/I$_X$ (f) between 10 successive one-minute acquisitions performed at delay $\tau$=170 ns. Measurements were all realised at T$_\mathrm{b}$=330 mK for a 1 $\mu$W mean laser excitation power.}
\end{figure}

The minimum acquisition time to ensure a sufficient signal to noise ratio constitutes one of our strongest experimental constraint. Indeed, indirect excitons are characterised by a long lifetime so that analysing spectrally their photoluminescence requires measurements typically 10 minutes long. Such elapsed time naturally challenges the experimental stability due to spectral diffusion. To quantify the strength of inhomogeneous broadening, we then computed the spectral width obtained by accumulating an ensemble of 10 successive one-minute acquisitions. As shown in Fig. 2.b, this leads to $\Gamma_\mathrm{X}$$\sim$  1.1 meV for $\tau\gtrsim$ 70 ns. For the same experiments, the value obtained by statistically averaging $\Gamma_\mathrm{X}$ from each individual realisation remarkably yields a spectral narrowing of about 0.4 meV, so that the spectral width reduces to 700 $\mu$eV (Fig.2.b). This difference directly manifests that our measurements suffer from fluctuations of the electrostatic environment, induced by a fluctuating trapping potential and/or a varying density of excess free carriers. In fact, this limitation is not very surprising since the excitons electric dipole amounts to about 500 Debye, so that time averaged electrostatic noise inevitably leads to  inhomogeneous broadening. Thus, E$_\mathrm{X}$ varies by at most 500 $\mu$eV over successive acquisitions (Fig.2.e).

Fig. 2.d illustrates further the spectral diffusion by comparing two successive measurements performed under unchanged conditions. Two limit cases case are presented, namely a broad profile ($\sim$ 1.5 meV wide) and a narrow one dominated by a 400 $\mu$eV lorentzian line. The latter realisation shows the result of the most stable experimental conditions, i.e. the ones with a minimum level of inhomogeneous broadening. $\Gamma_\mathrm{X}$ is then close to its theoretical expectation \cite{Zim_08}, and we actually verified in successive experiments, where the acquisition time was reduced by 5-fold, that the photoluminescence is homogeneously broadened in this situation \cite{Anankine_2016_t_coh}.  From Fig.2.b-d we thus deduce that the average inhomogeneous broadening amounts to at most 500 $\mu$eV. At this level, Fig.2.f shows that the relative variation of the integrated intensity ($\Delta$I$_X$/I$_X$) does not exceed $\pm$10$\%$ over 10 successive one-minute acquisition. 
This dispersion possibly reflects a fluctuating density of excess carriers that would vary the photoluminescence yield \cite{Peyghambarian_84}. Nevertheless, the measurements displayed in Fig.2.f indicate that we evaluate the fraction of bright excitons with about 20$\%$ precision long after the laser excitation (170 ns for Fig.2.f). This exactness is crucial for our studies because it sets the accuracy at which the occupation of dark states is then extracted, as further discussed in the following.

\subsection{Exciton cooling to sub-Kelvin bath temperatures}

We now study spectroscopically the trapped gas as the bath temperature is lowered. For that purpose, we performed measurements in a restricted parameter space where the strength of the loading laser excitation and the trapping potential remained unchanged. Accordingly, experimental conditions were well defined, but in turn the range of bath temperatures had to be limited to 0.33 $\leq$T$_\mathrm{b}\leq$ 3.5 K.

\begin{figure}[h!]\label{fig3}
\centerline{\includegraphics[width=.5\textwidth]{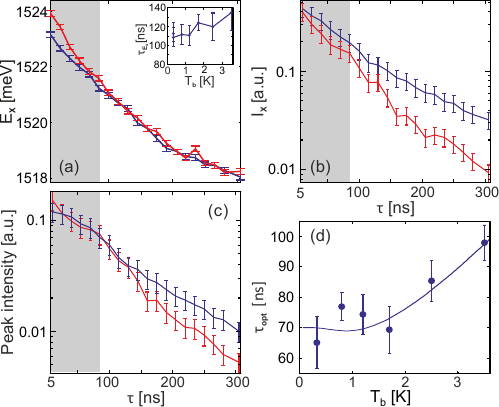}}
\caption{(a) Energy of the photoluminescence (E$_\mathrm{X}$) at T$_\mathrm{b}$=330 mK (red) and 2.5K (blue). (b) Decay of the photoluminescence integrated intensity (I$_\mathrm{X}$). (c) Dynamics of the maximum of the photoluminescence spectrum. (d) Optical decay time $\tau_\mathrm{opt}$ of the photoluminescence integrated intensity, versus T$_\mathrm{b}$. The solid line shows the result of the model detailed in Ref. \cite{Ivanov_2000}, assuming a radiative lifetime of 35 ns and a degeneracy temperature equal to 1.5 K. In (a) the inset shows the decay time of E$_\mathrm{X}$ as a function of T$_\mathrm{b}$. Experiments correspond to 3 days of continuous data acquisition, and the presented data have all been analyzed in a 3 $\mu$m wide region at the center of the trap.}
\end{figure}

In Figure 3.a we first display the dynamics of E$_\mathrm{X}$ from which the total density of excitons is extracted. Remarkably, for 100$\lesssim\tau\lesssim$300 ns, which corresponds to n$_X\lesssim$ 4 10$^{10}$ cm$^{-2}$, E$_\mathrm{X}$ varies by less than 500 $\mu$eV between 0.33 and 2.5 K. Given the stability of the trapping potential, we are led to conclude that the exciton density does not vary with T$_\mathrm{b}$ for this range of delays to the termination of the loading laser pulse. This is not very surprising since we expect that the exciton density is mostly given by the strength of the loading laser pulse, kept constant between these measurements. In fact, E$_\mathrm{X}$ only exhibits a slight dependence with T$_\mathrm{b}$ at short delays to the termination of the laser excitation (grey region in Figure 3), when the transient photocurrent is not fully evacuated. Afterwards, its dynamics can be approximated by a mono-exponential decay. The inset in Figure 3.a shows the extracted time constants $\tau_{E_X}$: for each bath temperature the precision of the fitting routine is reasonable and we deduce that  $\tau_{E_X}\sim$ 110 ns throughout the explored temperature range.

Unlike the total density, the occupation of the sole bright states strongly depends on T$_\mathrm{b}$. This behaviour is shown in Figure 3.b that presents  the dynamics of the integrated intensity I$_\mathrm{X}$ at T$_\mathrm{b}$=0.33 and 2.5 K. The transient regime (grey region) constitutes the only region where I$_\mathrm{X}$ varies weakly with T$_\mathrm{b}$. There, the population of low energy bright excitons is not controlled by the bath temperature but rather by the transient photocurrent, otherwise the population in the lowest energy bright states would increase between 2.5 and 0.33 K, and so would I$_\mathrm{X}$. Beyond the transient regime, Fig. 3.b clearly shows  that the photoluminescence intensity drops faster as the bath temperature is lowered. In an attempt to quantify this decrease, we modelled the decay of I$_\mathrm{X}$ with a single exponential after the transient regime, i.e., for $\tau\geq$80 ns. Figure 3.d shows that the resulting decay times $\tau_\mathrm{opt}$ drops from $\sim$ 100 ns at T$_\mathrm{b}$=3.5 K to $\sim$ 70 ns at 1.5 K, $\tau_\mathrm{opt}$ being rather constant at lower temperatures within the precision of our analysis ($\sim$10 ns). 

The optical decay time is expected to decrease with the exciton temperature \cite{Littlewood,Ivanov_2000,Andreani_90}. Indeed, only coldest bright excitons, with a kinetic energy lower than about 150 $\mu$eV ($\sim$ 1.5K), contribute to the photoluminescence (Fig.1.d). As a result, bright excitons decay faster as one lowers their temperature, simply because in average they "spend" more time in the radiative region of the energy band. The temperature dependence of the optical decay time thus provides direct information about the exciton statistics. Indeed, it is expected theoretically that Bose-Einstein statistics influences $\tau_\mathrm{opt}$ for both degenerate and non-degenerate gases \cite{Ivanov_2000}. The latter regime is found at high temperatures where Bose-Einstein statistics corrects the classically expected  linear dependence of the optical decay time with temperature \cite{Andreani_90}. For a degenerate gas,  $\tau_\mathrm{opt}$ then saturates to a value equal to twice the radiative lifetime \cite{Ivanov_2000}. Figure 3.d shows that such theoretical description reproduces well the measured variation of $\tau_\mathrm{opt}$, only assuming a degeneracy temperature equal to 1.5K and a radiative lifetime equal to 35 ns, which is reasonable for our heterostructure \cite{Mujlarov_2012}. The obtained agreement provides then a good indication that we experimentally study a Bose gas of indirect excitons which effective temperature is decreased by lowering the bath temperature.

\begin{figure}\label{fig4}
\centerline{\includegraphics[width=.5\textwidth]{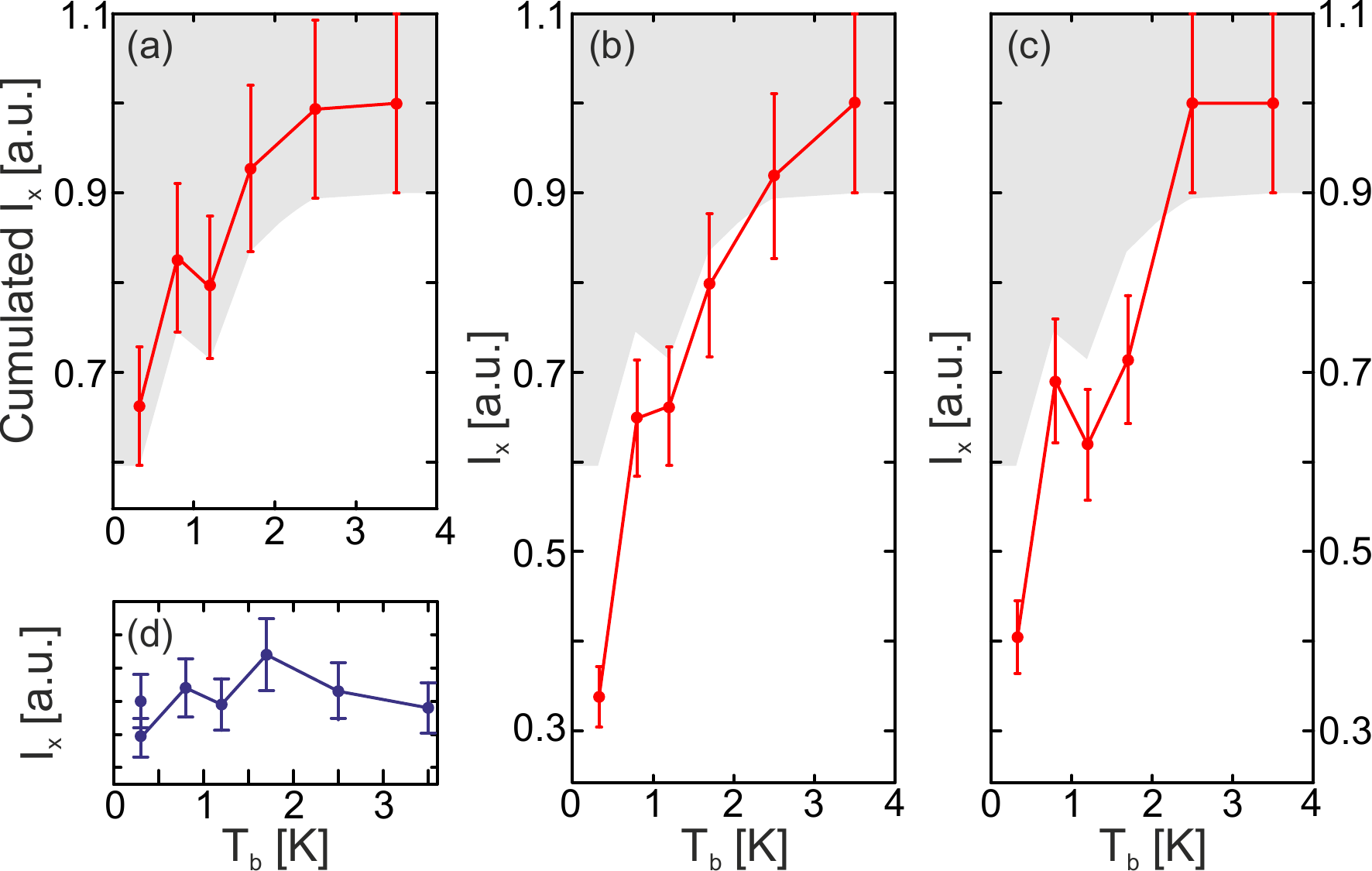}}
\caption{(a) Temperature variation of the cumulated integrated intensity, evaluated by integrating I$_X$ from $\tau$=80 to 350 ns, so that it is not affected by the transient photocurrent marked by the gray region in Fig.3. (b) Integrated intensity measured at $\tau$=190 ns, i.e., for $n_\mathrm{x}\sim$ 2 10$^{10}$ cm$^{-2}$, at $\tau$=250 ns (c) and  $\tau$=5 ns (d). In (b-c) the grey region displays the maximum error of our measurements deduced from panel (a), i.e. the region where the photoluminescence darkening can not be attributed to excitons dark-state condensation but instead to non-radiative losses.}
\end{figure}

\subsection{Quantum darkening: dark-state condensation}

The measurements summarised in Fig. 3 first show that above a few Kelvin the overall exciton population and its bright component both decay at a comparable rate  ($\tau_\mathrm{opt}\sim\tau_{E_X}\simeq$ 100 ns at T$_\mathrm{b}$=2.5K). We then deduce that the occupation of bright and dark states are similar in this regime which is then classical. On the other hand, $\tau_{E_X}$ is barely modified when T$_\mathrm{b}$ is lowered to sub-Kelvin temperatures whereas $\tau_\mathrm{opt}$ is decreased to about 70 ns at T$_\mathrm{b}$=330mK. These combined behaviours mark that the total exciton density has not varied with the bath temperature while the population of bright excitons is divided by at least 3-fold between 2.5K and 330mK (for $\tau\sim$ 190 ns, Fig. 3.b shows that I$_X$ is 3 times smaller at T$_\mathrm{b}$=330 mK than at 2.5 K). Since the total exciton density is conserved in the trap, $E_X$ being constant, the population of dark excitons has to be increased proportionally, resulting \textit{a priori} in a strong (quantum) imbalance between the occupations of bright and dark states at T$_\mathrm{b}$=330mK. Let us then note that such a dominant population of dark excitons is possibly built directly from bright excitons. Indeed, the exchange of the fermionic constituents (electron or hole) between two opposite-spin bright excitons convert these into opposite-spin dark ones \cite{Monique_dark_BEC}, thus acting as an effective channel for Bose stimulation.

We need to quantify the maximum error for the evaluation of the bright excitons density in order to show unambiguously that the photoluminescence quenching reveals an increased population of dark excitons. For that purpose, we computed the photoluminescence cumulated integrated intensity which adds up all the photons detected after the loading laser pulse. We thus evaluate the total number of bright excitons that have been confined in the trap after the loading phase. Let us stress that the cumulated intensity then also provides the total number of excitons that have been confined, regardless their brightness, otherwise non-radiative losses would play a significant role. Indeed, delocalised indirect excitons exhibit a spin coherence time which does not exceed a few nanoseconds \cite{High_2013,Beian_2015} while their spin relaxation time is typically bound to a few 10 ns \cite{Beian_2015,Holleitner_2010,Leonard_spin,Violante_2015,Andreakou_2015}. As a result, indirect excitons experience multiple scatterings between optically bright and dark manifolds during their lifetime ($\tau_{E_X}$), such that they can not remain trapped in dark states but instead must end by decaying radiatively. 


The cumulated integrated intensity as a function of the bath temperature is shown in Fig. 4.a,  which highlights a decrease by about 35$\%$ between 3.5 and 0.33 Kelvins. We then conclude that non-radiative losses can only account for up to 35$\%$ of the photoluminescence darkening,  by embracing every experimental limitation, such as the processes varying the rate of excitons radiative recombination in a temperature dependent manner, but also tunnelling processes that can lead to carrier escape from the trap. Altogether, experimental imperfections are then not sufficient to interpret our observations. Indeed, Figure 4.b.c shows that the amplitude of the photoluminescence darkening is well beyond this instrumental precision below a critical temperature of 1K. At T$_\mathrm{b}$=330 mK, it reaches approximately 70$\%$ 190 ns after the termination of the laser excitation when n$_X$$\sim$2 10$^{10}$ cm$^{-2}$, and about 60$\%$ 60 ns later when n$_X$$\sim$0.5 10$^{10}$ cm$^{-2}$  (outside of this delay range the darkening is not observed unambiguously). This confirms quantitatively that bright excitons are indeed converted into dark ones to conserve the total density at sub-Kelvin temperatures. Dark excitons then constitute about 70$\%$ of the population at  T$_\mathrm{b}$=330 mK. This marks a BEC since the energy splitting between optically bright and dark states only amounts to a few $\mu$eV \cite{Blackwood_1998,Mashkov_97}, that is 5 to 10-fold less than the thermal energy at 330mK. Finally, note that no darkening is observed shortly after the laser excitation, as expected since the exciton gas is then perturbed by excess carriers and probably not fully thermalised (Fig.4.d).

To further confirm that the photoluminescence quenching is due to dark-state condensation, we performed a control experiment where the integrated intensity I$_X$ was measured while the total exciton density n$_X$ was increased through the power of the loading laser P$_\mathrm{ex}$. Furthermore, the delay to the loading laser pulse was set to $\tau$=190ns, i.e. as for the measurements shown in Fig.4.b. After an initial transient regime at very low excitation  power (P$_\mathrm{ex}\lesssim$0.3$\mu$W) where the electrostatic confinement is probably unstable, Figure 5.a  shows that at T$_\mathrm{b}$=330 mK  I$_X$ increases linearly with P$_\mathrm{ex}$, for 0.4$\leq$P$_\mathrm{ex}\leq$0.8 $\mu$W and P$_\mathrm{ex}\geq$2 $\mu$W. Since we verified for these experiments that the total exciton density varies linearly with the laser power (Fig.5.b), the trapped gas behaves classically in this regime. Indeed, for a classical gas bright and dark states have similar populations so that the occupation of bright states has to increase proportionally to the loading laser power. By contrast, we recover the quantum regime in the red region of Fig.5.a (0.8$\leq$P$_\mathrm{ex}\leq$1.5 $\mu$W) where the photoluminescence intensity is quenched while the total exciton density increases. We then note that in Fig.4.b this quenching is resolved for the same delay $\tau$=190 ns but by varying the bath temperature, for a total exciton density fixed by P$_\mathrm{ex}$=1 $\mu$W. Thus, in Fig.5.a we access the threshold density n$_X^{(c)}$ for the dark-state condensation in the trap, n$_X^{(c)}\sim$1.7 10$^{10}$ cm$^{-2}$ at T$_\mathrm{b}$=330mK, whereas in Fig.4.b we measure the critical temperature for the condensation, T$_c\sim$1K for n$_X$$\sim$2 10$^{10}$ cm$^{-2}$. Finally, we verify in Figure 5.a that the photoluminescence quenching is not observed for T$_\mathrm{b}\geq$1.3K.

\begin{figure}\label{fig5}
\centerline{\includegraphics[width=.5\textwidth]{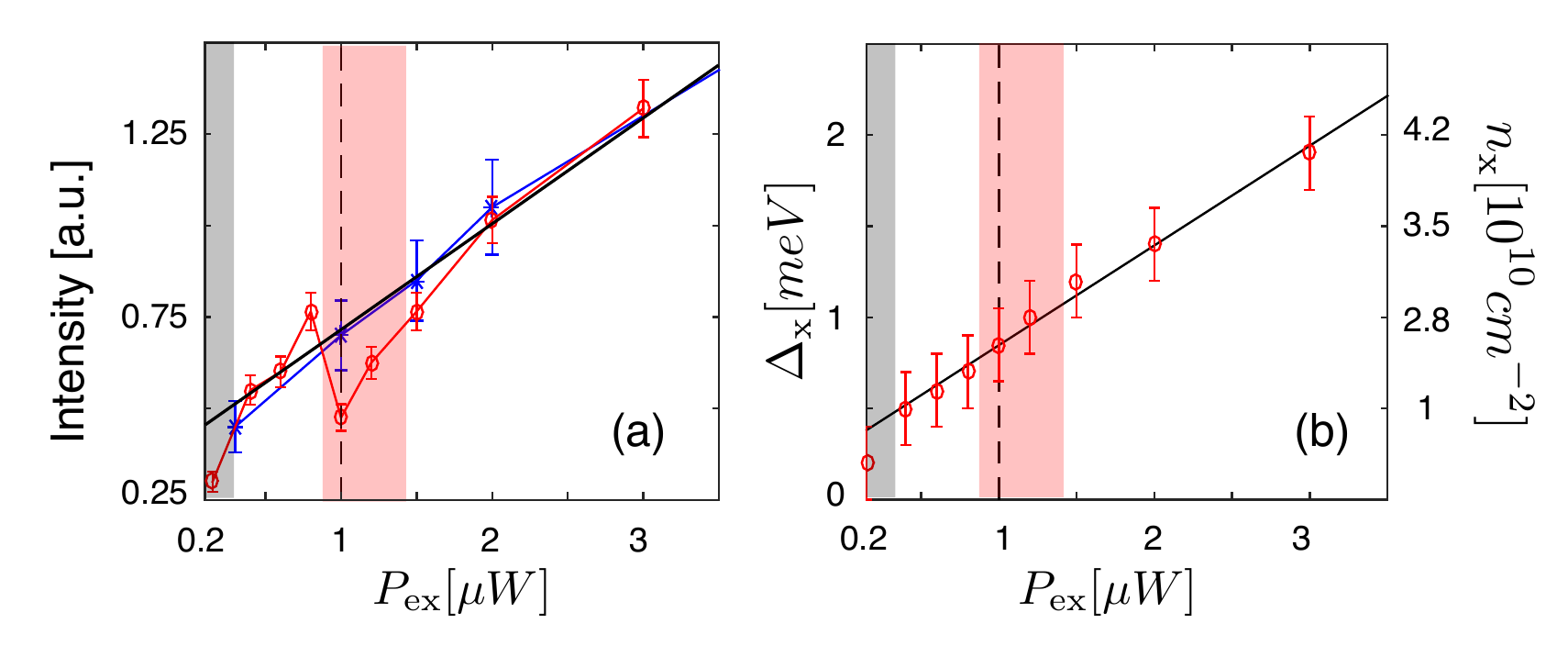}}
\caption{(a) Integrated intensity I$_\mathrm{X}$ as a function of the power of the loading laser pulse P$_\mathrm{ex}$, at T$_\mathrm{b}$=330K (red circles) and 1.3K (blue stars). The gray region marks the regime where the excitons electrostatic environment is unstable, the solid black line shows a linear regression of I$_X$, i.e. its classically expectation. (b) Blue-shift of the photoluminescence energy $\Delta E_X$ and total exciton density n$_X$ deduced from the model described in Ref. \cite{Ivanov_2010}. The solid line displays the expected linear increase. Experiments have been realized for a delay to the loading laser pulse set to $\tau$=190 ns. The vertical dashed lines indicate the value of P$_\mathrm{ex}$ used for the experiments discussed in Fig.2 to Fig.4.}
\end{figure}

In Fig. 5.a, let us compare the integrated intensity at T$_\mathrm{b}$=330mK to its classical expectation (black line) for P$_\mathrm{ex}$=1 $\mu$W. Thus, we deduce that I$_X$ is reduced by 30$\%$ in the condensed phase. Importantly, this amplitude agrees closely with the one we can obtain from Fig.4.b where we also have set P$_\mathrm{ex}$=1 $\mu$W. Indeed, assuming that excitons are classically distributed at T$_\mathrm{b}$=3.5K since no darkening is found in this regime, by taking into account non-radiative losses we deduce from Fig.4.b that about 70$\%$ of the exciton population is condensed in dark states at T$_\mathrm{b}$=330 mK. Moreover, the red region in Fig. 5.a corresponds to a total density 1.7 10$^{10}$ cm$^{-2}$ $\lesssim n_\mathrm{x}\lesssim$ 2.7 10$^{10}$ cm$^{-2}$(Fig.5.b). This restricted range of densities for the photoluminescence quenching does not contradict the one we deduced by analyzing the photoluminescence dynamics shown in Figure 4.b-c.

 \subsection{Conclusions}

We would like to point out that our experiments, relying on time and spatially resolved spectroscopy, can only evidence the dark-state condensation through a loss of photoluminescence intensity for a calibrated total exciton density. To unambiguously relate this loss to quantum condensation, we have verified that non-radiative channels can not account for our observations. This point is crucial to reach clear conclusions, however it had never been verified before to the best of our knowledge, including in studies where anomalously weak photoluminescence emissions had been discussed \cite{Alloing_2014,Rapaport_2013,Rapaport_2016,Combescot_ROPP}. Moreover, the photoluminescence quenching only occurs for a very narrow range of densities in the dilute regime (Figures 4 and 5),  contrasting with recent works that reported a dark liquid of indirect excitons \cite{Rapaport_2013,Rapaport_2016}.

According to theoretical predictions \cite{Combescot_2012,Combescot_2015}, and also to recent experiments that we performed onto the same heterostructure \cite{Anankine_2016}, in the regime of photoluminescence quenching the condensate is not only made of dark excitons but instead consists of a dominant dark component coherently coupled to a weaker bright one. This coupling between bright and dark states is the result of fermion exchanges between excitons, which favour the introduction of a bright component to the quantum phase at experimentally studied densities \cite{Combescot_2012}. The dark condensate then becomes "gray" and is possibly studied through its weak coherent photoluminescence. It is this emission that has allowed us to irrefutably verify the conclusions we reach here, through experiments where we have reported macroscopic spatial coherence, and quantised vortices revealing the long-sought superfluidity of excitons \cite{Anankine_2016}. Let us then stress that observing exciton superfluidity required to identify the regime of photoluminescence quenching quantified here. In fact, this behaviour constitutes the key to access collective quantum states.

\textbf{Acknowledgments:} The authors are grateful to Kamel Merghem and to Suzanne Dang for assistance, as well as to Monique Combescot and Roland Combescot for a critical reading of the manuscript and for many enlightening discussions. This work has been supported by the EU FP7-ITN INDEX, EU FP7-CIG X-BEC, by OBELIX from the french Agency for Research (ANR-15-CE30-0020), by MINECO (SEVERO OCHOA Grant SEV-2015-0522, FISICATEAMO FIS2016-79508-P),  by the Generalitat de Catalunya (SGR 874 and CERCA program) and by the Fundacio Privada Cellex.

\end{document}